\documentstyle[12pt]{article}
\begin{document}

\baselineskip19pt

\begin{center}
{\bf {\Large Elementary constructive approach to the higher-rank
numerical ranges of unitary matrices.}} \vspace{0.3cm}

{\bf A.Ya.Kazakov } \vspace{0.5cm} \\Laboratory of quantum information, \\%
St.-Petersburg state university of aerospace instrumentation,

67 B.Morskaya Str., St.-Petersburg, 190000 Russia \vspace{0.2cm}

\vspace{0.5cm}
\end{center}

\vspace{0.5cm}

\begin{abstract}
Some problems of the quantum error-correcting codes theory can be
reduced to the investigation of the higher-rank numerical ranges
of the operators related to the error operators. We constructively
verify a conjecture on the structure of higher-rank numerical
range for unitary matrices.
\end{abstract}

\section{Introduction}

Quantum error correction is one of the main directions in the
developing of the quantum information theory since middle of the
1990, see \cite{Shor}- \cite{KL}. Recently in series of papers
\cite{CKZ} - \cite{CHKZ} it was introduced one approach to the
realization of error-correcting codes for quantum channels.
Results of \cite{CKZ} - \cite{CHKZ} give the possibility to reduce
the realization of the correctable codes to the matrix analysis
problem, namely, to the study of the ''higher-rank numerical
range'' of the operators related to the ''error operator'' of the
quantum channel. The ''higher-rank numerical range'' is a
generalization of the usual notion of the operator spectrum.
Namely, let $\bf{H}$ be a finite-dimensional Hilbert space,
$\bf{B(H)}$ be a set of operators
acting on $\bf{H}$, $\sigma \in \bf{B(H)}$. For $k\geq 1$ the \textit{%
rank-k numerical range } of $\sigma $ is the subset of complex
plane $\Lambda _k( \sigma ) =\left\{ {\lambda \in {\bf{C}} : P\sigma %
P=\lambda P }\right. $ for some k-dimensional orthogonal
projections $P$ on $\left. {\bf{H}}\right\} $. The key problem is
a description of $\Lambda _k( \sigma ) $ for a given operator
$\sigma $ in explicit terms. The next statement was proved in
\cite {CHKZ}.

{\bf Proposition 1.}  Let $\bf{H}$ be an $N-$dimensional Hilbert
space, $k\geq 1$ be a positive integer, $\sigma $ be a normal
matrix, then
\begin{equation}
\Lambda _k\left( \sigma \right) \subseteq \Omega _k(\sigma ),  \label{ff1}
\end{equation}
 where
\[
\Omega _k(\sigma )=\cap conv(\Gamma ),
\]
 and $\Gamma $ runs through all $(N-k+1)-$point subsets (counting
multiplicities) of the set of eigenvalues $spec(\sigma )$ for $\sigma ,$ $%
conv(\Gamma )$ means the convex hulls of the set $\Gamma $.

It was conjectured in \cite{CKZ}, that the conversion of  this
statement is valid.

{\bf Conjecture.} For the normal matrix $\sigma $
\begin{equation}
 \Lambda _k\left( \sigma \right)
=\Omega _k(\sigma ).  \label{ff2}
\end{equation}

For the brevity we will denote Conjecture for given $N,k$ as
$(N,k)$. This statement is not proved in general case, some
particular cases were discussed in \cite{CHKZ}.  In particular,
the next propositions were checked in \cite{CHKZ}

{\bf Proposition 2.}  Conjecture holds if and only if the
corresponding statement holds for all unitary matrices.

{\bf Proposition 3.} Conjecture (\ref{ff2}) is valid for $N\geq
3k,$ $k\geq 2,$ $(5,2),$ $(8,3),$ and, generally, $(3k-1,k),k\geq
2.$

Conjecture $(N,k), N\geq 3k$, was verified in \cite{CHKZ}
explicitly. The corresponding construction was presented with help
of simple and elementary terms, see discussion below. However, the
verification of the Conjecture $(5,2),$ $(8,3),$ and $(3k-1,k)$
was given \cite{CHKZ}  non-constructively. But at the realization
of the quantum error-correcting codes it is necessary to get an
explicit description of the corresponding objects, such as
projector $P$. Note, that full proof of the Conjecture was
obtained in \cite{LS} with help of more advanced technique.

The aim of this note is to modify an elementary approach of the
\cite{CHKZ} and to suggest a constructive verification of the
Conjecture $(3k-1,k),k\geq 2$ and $(3k-2,k),k\geq 5$. We will
consider here the mathematical details only, initial motivation
and discussion of possible applications in the theory of quantum
error-correcting codes can be found in \cite{CKZ} - \cite{CHKZ}.

\section{General considerations}

As it follows from Proposition 2, we can discuss a unitary matrix
$\sigma $, so its spectrum belongs to the unit circle. Let
eigenvalues of $\sigma $
are $\lambda _j=\exp \left( i\theta _j\right) ,$ $j=1,2,...,N,$ such that $%
0\leq \theta _1\leq \theta _2\leq ...\theta _N<2\pi .$ We extend the
numbering of the $\lambda _j$ and $\mid \psi _j>$ cyclically if it is
necessary. For multiple eigenvalues the numbering is arbitrary, and we
choose an orthonormal system of eigenvectors $\mid \psi _j>\in \bf{H},$%
\begin{equation}
\sigma \mid \psi _j>=\lambda _j\mid \psi _j>,j=1,2,...,N.  \label{ff3}
\end{equation}
Let $\lambda \in \Lambda _k\left( \sigma \right) $, it means, that
k-dimensional orthogonal projection $P$ exist, for which
\begin{equation}
P\sigma P=\lambda P.  \label{kl1}
\end{equation}
Let
\begin{equation}
P=\sum_{s=1}^k\mid \varphi _s><\varphi _s\mid   \label{kl2}
\end{equation}
for some set of orthonormal vectors $\left\{ \mid \varphi _1>,\mid \varphi
_2>,...,\mid \varphi _k>\right\} $ and
\begin{equation}
\mid \varphi _s>=\sum_mz_{sm}\mid \psi _m>,  \label{kl21}
\end{equation}
then normalization means, that
\begin{equation}
\sum_m\mid z_{sm}\mid ^2=1,  \label{kl31}
\end{equation}
and orthogonality means, that
\begin{equation}
\sum_mz_{sm}\overline{z_{pm}}=0,s\neq p.  \label{kl3}
\end{equation}
Relation (\ref{kl1}) reads in our notations:
\begin{equation}
\sum_m\lambda _m\mid z_{sm}\mid ^2=\lambda ,s=1,2,...,k.  \label{kl4}
\end{equation}
For the convenience of following discussions we formulate the inversion of
these considerations as a proposition.

{\bf Proposition 4.}  If for given $\lambda $ we can find a set of
vectors $\left\{ \mid \varphi
_1>,\mid \varphi _2>,...,\mid \varphi _k>\right\} $ which satisfy relations (%
\ref{kl21})- (\ref{kl4}), then $\lambda \in \Lambda _k\left(
\sigma \right) $ and corresponding projector $P$ is described by
relation (\ref{kl2}) .

First of all we cite here one more result of \cite{CHKZ} which is useful in
what follows.

{\bf Proposition 5.} Given integers $i,j$ with $i<j<i+N,$ let
$D(i,j)$ denote the convex subset of $\bf{C}$ bounded by the line
segment from $\lambda _i$ to $\lambda _j$ and the counterclockwise
circular arc from $\lambda _j$ to $\lambda _i$. Then
\begin{equation}
\Omega _k(\sigma )=\cap _{i=1}^ND(i,i+k).  \label{ff4}
\end{equation}

The next simple result which will be exploited below follows from
Proposition 5.

{\bf Corollary.} Let $T(\lambda _i, \lambda _j, \lambda _m)$,
$i<j<m$, is a triangle with vertexes $\left\{(\lambda
_i)\right\},\left\{(\lambda _j)\right\},\left\{(\lambda
_m)\right\}$. If $\mid j-i \mid <k$,$\mid m-j \mid <k$,$\mid N+i-m
\mid <k$, then $\Omega _k (\sigma )\subset T(\lambda _i, \lambda
_j, \lambda _m)$, so for any $\lambda \in \Omega _k (\sigma )$
exist nonnegative numbers $p_i,p_j,p_m$, for which the following
relations are valid:
\[
p_i+p_j+p_m=1,
\]
\[
\lambda _ip_i+\lambda _jp_j+\lambda _mp_m=\lambda .
\]
Note, that these relations are a special case of relations
(\ref{kl31}) and (\ref{kl4}). We will associate with such triangle
a normalized vector
\begin{equation}
 \mid \varphi _s>=\sqrt{p_i}\mid \psi _i>+\sqrt{p_j}\mid \psi
_j>+\sqrt{p_m}\mid \psi _m>. \label{vect}
\end{equation}

 As was mentioned above, Conjecture $(N,k), N \geq 3k,$ was proved
 in \cite{CHKZ}. Namely, corresponding procedure includes a
 construction of $k$ triangles satisfying conditions of Corollary.
 These triangles have not common vertexes, so corresponding vectors
 (\ref{vect}) are orthogonal each other. Full set of these vectors
 satisfy Proposition 4.

Here we present some modification of an elementary approach
\cite{CHKZ}. Namely, we use a set of $k$ triangles too, but we
permit existence of common vertex either for one pair of triangles
(Conjecture $((3k-1),k)$ or for two pairs of triangles (Conjecture
$((3k-2),k)$. The key result in our considerations is the
following statement.

{\bf Proposition 6.} Let we have two set of numbers $p_t,q_r\geq
0,$ $t\in T\subset \left\{ 1,2,...,N\right\} ,$ $r\in R\subset
\left\{ 1,2,...,N\right\} ,$ $T\cap R=\emptyset ,$ which satisfy
the next conditions:
\begin{equation}
p_1+\sum_{t\in T}p_t=1  \label{n1}
\end{equation}
\begin{equation}
q_1+\sum_{r\in R}q_r=1  \label{n2}
\end{equation}
\begin{equation}
\lambda _1p_1+\sum_{t\in T}\lambda _tp_t=\lambda ,  \label{res1}
\end{equation}
\begin{equation}
\lambda _1q_1+\sum_{r\in R}\lambda _rq_r=\lambda .  \label{res2}
\end{equation}
 If either $p_1\leq 1/2$ or $q_1\leq 1/2,$ then there are two
orthonormal vectors
\begin{equation}
\mid \varphi _1>=z_{11}\mid \psi _1>+\sum_{s\in T\cup R}z_{1s}\mid \psi _s>,
\label{res3}
\end{equation}
\begin{equation}
\mid \varphi _2>=z_{21}\mid \psi _1>+\sum_{s\in T\cup R}z_{2s}\mid \psi _s>,
\label{res4}
\end{equation}
satisfying relations (\ref{kl31})-(\ref{kl4}).

Proof. Let
\begin{equation}
\mid \varphi _1>=\left[ \sqrt{p_1}\cos \theta +i\sqrt{q_1}\sin \theta
\right] \mid \psi _1>+\exp \left( i\alpha \right) \cos \theta \sum_t\sqrt{p_t%
}\mid \psi _t>+\exp \left( i\beta \right) \sin \theta \sum_r\sqrt{q_r}\mid
\psi _r>,  \label{res5}
\end{equation}
\begin{equation}
\mid \varphi _2>=\left[ \sqrt{p_1}\cos \tau +i\sqrt{q_1}\sin \tau \right]
\mid \psi _1>+\cos \tau \sum_t\sqrt{p_t}\mid \psi _t>+\sin \tau \sum_r\sqrt{%
q_r}\mid \psi _r>.  \label{res6}
\end{equation}
Simple calculations with help of (\ref{n1})-(\ref{res2}) confirm,
that relations (\ref{kl31}) and (\ref{kl4}) are valid for these
vectors. We have to find values $\theta ,\tau ,\alpha ,\beta $ in
 order to get orthogonality of vectors $\mid \varphi _1>,\mid
\varphi _2>$. We obtain the following condition:
\[
<\varphi _1,\varphi _2>=p_1\cos \theta \cos \tau +q_1\sin \theta \sin \tau
+\exp \left( i\alpha \right) \cos \theta \cos \tau \left( 1-p_1\right) +\exp
\left( i\beta \right) \sin \theta \sin \tau \left( 1-q_1\right) +
\]
\[
i\sqrt{p_1q_1}\left( \sin \theta \cos \tau -\cos \theta \sin \tau \right) =0,
\]
here we take into account relations (\ref{n1}) and (\ref{n2}). Let
$x=\tan \theta ,y=\tan \tau .$ Separating real and imaginary parts
of the last expression, one can get the following pair of
equations:
\begin{equation}
p_1+q_1xy+\cos \alpha \left( 1-p_1\right) +\cos \beta \left( 1-q_1\right)
xy=0,  \label{res7}
\end{equation}
\begin{equation}
\sqrt{p_1q_1}\left( x-y\right) +\sin \alpha \left( 1-p_1\right) +\sin \beta
\left( 1-q_1\right) xy=0.  \label{res8}
\end{equation}
Excluding $y$, we obtain quadratic equation for $x$:
\begin{equation}
x^2+Ax+B=0,  \label{res9}
\end{equation}
where
\begin{equation}
A=\frac{(p_1-1)\sin \alpha \left[ q_1+(1-q_1)\cos \beta \right] +(q_1-1)\sin
\beta \left[ (p_1-1)\cos \alpha -p_1\right] }{\sqrt{p_1q_1}\left(
q_1+(1-q_1)\cos \beta \right) },  \label{res10}
\end{equation}
\begin{equation}
B=\frac{p_1+(1-p_1)\cos \alpha }{q_1+(1-q_1)\cos \beta }.  \label{res11}
\end{equation}
Equation (\ref{res9}) has real root, if the next condition holds:
\[
A^2\geq 4B,
\]
or, in more details,
\begin{equation}
\left\{ (p_1-1)\sin \alpha \left[ q_1+(1-q_1)\cos \beta \right] +(q_1-1)\sin
\beta \left[ (p_1-1)\cos \alpha -p_1\right] \right\} ^2\geq   \label{res12}
\end{equation}
\[
4p_1q_1\left[ p_1+(1-p_1)\cos \alpha \right] \left[ q_1+(1-q_1)\cos \beta
\right] .
\]
Note, that if either $p_1\leq 1/2$ or $q_1\leq 1/2$ we can get
non-positive right-hand side of the last expression by the
corresponding choice of the parameters $\alpha ,\beta $. Then
condition (\ref {res12}) holds, we can calculate corresponding
(real) values $x,y$ or, in other words, $\theta ,\tau $ and
construct the pair of orthonormal vectors $\mid \varphi _1>,\mid
\varphi _2>$ in explicit terms. The result follows.

{\bf Definition }. Let $p_t\geq 0, t \in T\subset \left\{ {
1,2,...,N}\right\}   $, and for given $\lambda $
\begin{equation}
p_{t_1}+\sum_{t\neq t_1}p_t=1  \label{nt1}
\end{equation}
\begin{equation}
\lambda _{t_1}p_{t_1}+\sum_{t\neq t_1}\lambda _tp_t=\lambda ,
\label{rest1}
\end{equation}
and $p_{t_1} \leq 1/2$. We call the  point $\left\{ t_1 \right\}$
{\it weak vertex} of the polygon generated by $T$.

In what follows we will construct $k$ triangles $T\left\{\lambda
_i, \lambda _j, \lambda _m\right\}$, each of them will satisfy
condition of the Corollary. Note, that each such triangle contains
two weak vertexes. For the Conjecture $(3k-1,k)$ only one pair of
triangles (only two pairs for the Conjecture $(3k-2,k)$) will have
one common  vertex, weak for one of triangles. So, we can apply
Proposition 6 and get a pair of necessary vectors $\mid \varphi
_1>, \mid \varphi _2>$ (two pairs for Conjecture $(3k-2,k)$,
respectively). For the remaining $(k-2)$ ($(k-4)$ for Conjecture
$(3k-2,k)$) triangles corresponding vectors will be defined by
relation (\ref{vect}). These vectors will be normalized and
orthogonal each other, and this set of vectors will satisfy
conditions (\ref{kl31})-(\ref{kl4}).

\section{Constructive verification of the Conjecture  $(3k-1,k).$}
So, our aim is to find a necessary system of triangles. In order
to clarify details, we firstly consider $N=5,k=2.$ The spectrum of
the unitary operator $\sigma $ is depicted on figure 1. Note, that
some eigenvalues can coincide, but we represent them as different
points for more clearness. Let $\lambda \in \Omega _2(\sigma ).$
In accordance with Proposition 2,
\[
\lambda \in T\left\{ 1,3,5\right\} \cap T\left\{ 1,2,4\right\}
\cap T\left\{ 2,4,5\right\}.
\]
 As a first triangle, appearing in Proposition 6,
we take $T\left\{ 1,3,5\right\} $ . Note, that either vertex
$\left\{1\right\}$ or vertex $\left\{5\right\}$ is weak vertex of
this triangle. In the first case we take $T\left\{ 1,2,4\right\} $
as a second triangle appearing in Proposition 6. In this case
$T\left\{ 1,3,5\right\} \cap T\left\{ 1,2,4\right\} =\left\{
1\right\} .$ If the vertex $\left\{5\right\}$ is weak vertex of
triangle $T\left\{ 1,3,5\right\} $, we take $T\left\{
2,4,5\right\} $ as a second triangle,  $T\left\{ 1,3,5\right\}
\cap T\left\{ 2,4,5\right\} =\left\{ 5\right\} $. In both
situations intersection of chosen triangles contains only one
vertex, which is weak for triangle $T\left\{ 1,3,5\right\} $ and
we can apply Proposition 6. Then we obtain the pair of vectors
satisfying relations (\ref{kl31})-(\ref{kl4}).

As the second example we consider $N=3k-1,k\geq 3$. The spectrum
in this situation is depicted on figure 2. If $\lambda \in \Omega
_k(\sigma ),$ then, in accordance with Corollary, $\lambda $
belongs to the triangle $T\left\{ 1,k+1,2k+1\right\} .$ In that
triangle either vertex $\left\{1\right\}$ or $\left\{2k+1\right\}$
is weak one. Let, for distinctness, it is $\left\{1\right\}$
(there is a symmetry of the picture). Then we take as the second
triangle $T\left\{ 1,k,2k\right\} $. The remaining $(k-2)$
triangles are $T\left\{ k-1,2k-1,3k-2\right\} ,T\left\{
k-2,2k-2,3k-3\right\} $ etc. As follows from the Corollary,
$\Omega _k(\sigma )$ belongs to intersection of all triangles.
Note, that only triangles
 $T\left\{ 1,k+1,2k+1\right\} ,$ $T\left\{ 1,k,2k\right\} $
 have one common vertex (which is weak
for the first triangle). Applying Proposition 6 to the triangles
$T\left\{ 1,k+1,2k+1\right\} ,$ $T\left\{ 1,k,2k\right\} $, we can
construct a pair of orthogonal vectors $\mid \varphi _k>,\mid
\varphi _{k-1}>$, which satisfy conditions (\ref{kl31}),
(\ref{kl4}). For each triangle $T\left\{ k-m,2k-m,3k-m-1\right\}
,m=1,2,...,k-2,$ we take associated by (\ref{vect}) vectors,
\[
\mid \varphi _m>=\sqrt{p_{k-m}}\mid \psi _{k-m}>+\sqrt{p_{2k-m}}\mid \psi
_{2k-m}>+\sqrt{p_{3k-m-1}}\mid \psi _{3k-m-1}>,
\]
where positive numbers $p_{k-m},p_{2k-m},p_{3k-m-1}$ are determined by
relation
\[
\lambda =\lambda _{k-m}p_{k-m}+\lambda _{2k-m}p_{2k-m}+\lambda
_{3k-m-1}p_{3k-m-1}.
\]
As was mentioned above, vectors $\left\{ \mid \varphi
_m>,m=1,2,...,k\right\} $ are orthogonal each other. So, we have
constructed the necessary set of vectors and the corresponding
orthogonal projector is given by relation (\ref{kl2}).

Note, that correctness of  Conjecture $((3k-1)m,km),$ $m\geq 2$
follows immediately from our results.

\section{Constructive verification of the Conjecture $(3k-2,k).$}

Now we consider  Conjecture $(3k-2,k), k \geq 5$. In order to
verify this situation we have twice apply Proposition 6 .

For the convenience we begin from Conjecture $(13,5)$ (see figure
3) . First of all we depict the triangle $T\left\{ 1,4,9\right\}
$. In this triangle either $\left\{ 1\right\} $ or $\left\{
4\right\} $ is a weak vertex. There is an evident symmetry of our
figure on this stage, and we choose  vertex $\left\{ 1\right\} $.
Then next triangle will be $T\left\{ 1,6,11\right\} $, which has
one common vertex with triangle $T\left\{ 1,4,9\right\} $. With
help of  Proposition 6 we can construct two orthonormal vectors
$\mid \varphi _1>,\mid \varphi _2>,$ satisfying relations
(\ref{kl31})-(\ref{kl4}). As the next triangle we choose $T\left\{
3,8,12\right\} $, where either $\left\{ 8\right\} $ or $\left\{
12\right\} $ is a weak vertex.

1) Let $\left\{ 8\right\} $ is the weak vertex of the triangle
$T\left\{ 3,8,12\right\} $. Then we take triangle $T\left\{
5,8,13\right\} $ and for pair of triangles $T\left\{
3,8,12\right\} ,T\left\{ 5,8,13\right\} $ we construct with help
of Proposition 6 the pair of orthogonal vectors $\mid \varphi
_3>,\mid \varphi _4>,$ satisfying relations
(\ref{kl31})-(\ref{kl4}). Residuary vertexes gives us the last
triangle $T\left\{ 2,7,10\right\} $, which generates the fifth
necessary vector in accordance with (\ref{vect}).

2)  Let $\left\{ 12\right\} $ is the weak vertex of the triangle
$T\left\{ 3,8,12\right\} $. Then we choose triangle $T\left\{
2,7,12\right\} $ and for pair of triangles $T\left\{
3,8,12\right\} ,T\left\{ 2,7,12\right\} $ we construct with help
of  Proposition 6 the pair of orthogonal vectors $\mid \varphi
_3>,\mid \varphi _4>,$ satisfying relations
(\ref{kl31})-(\ref{kl4}). Residuary vertexes gives us the last
triangle $T\left\{ 5,10,13\right\} $ and we obtain the fifth
vector, associated with this triangle by (\ref{vect}).

Let now consider Conjecture $(3k-2,k)$ for $k >5$, see figure 4.
First triangle is triangle $T\left\{ 1,k-1,2k-1\right\} $, and
either $\left\{ 1\right\} $ or $\left\{ k-1\right\} $ is a weak
vertex. Due to symmetry we can choose any of them, and we choose
$\left\{ 1\right\} $. The next triangle is $T\left\{
1,k+1,2k+1\right\} $, which has one common vertex with $T\left\{
1,k-1,2k-1\right\} $. So, proposition 6 gives the possibility to
construct a pair vectors $\mid \varphi _1>,\mid \varphi _2>$ with
necessary properties. Then we choose triangle $T\left\{
k-2,2k-2,3k-3\right\} $. Here either $\left\{ 2k-2\right\} $ or
$\left\{ 3k-3\right\} $ is a weak vertex.

1) Let $\left\{ 2k-2\right\} $ is a weak vertex. Then we choose as
a next triangle  $T\left\{ k,2k-2,3k-2\right\} $. With help of
Proposition 6 we construct one more pair of vectors $\mid \varphi
_3>,\mid \varphi _4>$ with necessary properties. The triangle
$T\left\{ 2,k+2,2k\right\} $ gives the fifth vector $\mid \varphi
_5>$. Additional $(k-5)$ vectors can be constructed with help of
$k-5$ triangles $T\left\{ m,k+m,2k+m-1\right\} , m=3,4,...,k-3.$
Each such triangle satisfies condition of Corollary and relation
(\ref{vect}) gives us corresponding vector.

2) Let now $\left\{ 3k-3\right\} $ is a weak vertex. As a next
triangle we choose $T\left\{ k-3,2k-3,3k-3\right\} $. Applying
Proposition 6 we construct one more pair of vectors $\mid \varphi
_3>,\mid \varphi _4>$ with necessary properties. The triangle
$T\left\{ k,2k,3k-2\right\} $ gives us the fifth vector $\mid
\varphi _5>$. Additional $(k-5)$ vectors can be constructed with
help of $(k-5)$ triangles $T\left\{ m,k+m,2k+m-1\right\} ,
m=2,3,...,k-4$ and relation (\ref{vect}).

Evidently, that Conjecture $((3k-2)s,ks), k \geq 5, s=2,3,...,$
follows from this result.

\section{Conclusion}
We have discussed the "higher-rank numerical ranges" method of
constructing error-correcting codes for quantum channels. The
realization of the correctable codes is reduced in this framework
to the matrix analysis problem, which was thoroughly considered in
papers \cite{CKZ} - \cite{CHKZ} and solved in \cite{LS}.
Realization of the error-correcting codes is based on explicit
description of higher-rank numerical ranges of operators related
to the error operators. Corresponding constructive description was
obtained in \cite{CHKZ} for $N\geq 3k$ in elementary terms. Here
we have presented some modification of this construction, which
can be applied for $N=3k-1, k\geq 2$ and $N=3k-2, k\geq 5$.  These
results can be useful at constructing error-correcting codes for a
special classes of quantum channels {CHKZ}.

\end{document}